\documentclass[conference,compsoc]{IEEEtran}
\usepackage[utf8]{inputenc}
\usepackage{graphicx}
\usepackage[%
    colorlinks=true,
    pdfborder={0 0 0},
    linkcolor=cyan
]{hyperref}
\usepackage{hyperref}
\hypersetup{
     colorlinks   = true,
     citecolor    = black
}
\usepackage{cite}
\usepackage{amsmath,amssymb,amsfonts}
\usepackage{algorithmic}
\usepackage{graphicx}
\usepackage{textcomp}
\usepackage{xcolor}
\usepackage{array}
\usepackage{caption}
\def\BibTeX{{\rm B\kern-.05em{\sc i\kern-.025em b}\kern-.08em T\kern-.1667em\lower.7ex\hbox{E}\kern-.125emX}}
\begin{document}    
\title{A systematic literature review on Ransomware attacks}
\author{\IEEEauthorblockN{Krishnaben Chiragkumar Bhavsar }
\IEEEauthorblockA{\textit{School of Engineering } \\
\textit{University of Guelph}\\
Guelph, Canada \\
E-mail: kbhavsar@uoguelph.ca}
\and
\IEEEauthorblockN{Nishtha Pareshkumar Patel}
\IEEEauthorblockA{\textit{School of Engineering} \\
\textit{University of Guelph}\\
Guelph, Canada \\
E-mail: nishthap@uoguelph.ca}
\and
\IEEEauthorblockN{Shweta Chiragkumar Vasoya }
\IEEEauthorblockA{\textit{School of Engineering } \\
\textit{University of Guelph}\\
Guelph, Canada \\
E-mail: svasoya@uoguelph.ca}
}

\maketitle

\begin{abstract}
In the area of information technology, cybersecurity is critical. Information security is one of today's highest priorities. Cyber attacks, which are on the rise and include Ransomware, are the first thing that springs to mind when we think about cybersecurity. To counteract cybercrime, several governments and companies employ a range of strategies. Despite several cybersecurity measures, ransomware continues to terrify people. The victim's files or data are encrypted by ransomware attackers, who then demand payment to decrypt the data. In order to safeguard the cyber environment, organisations, and user assets, a variety of tools, rules, security guards, security ideas, guidelines, risk management, activities, training, insurance, and best practises are utilised. Despite of this, attacks on computers and networks have risen exponentially in recent years. The most devastating cyberattack in the world is ransomware, which locks user devices' interfaces or encrypts user files to prevent users from using their devices. Typically, a Trojan that appears to be a legitimate file and is delivered as an email attachment tricks the victim into downloading or opening it, which is how ransomware assaults are conducted. One well-known example, the WannaCry worm, on the other hand, automatically switched across systems without any human input. This paper examines ransomware assaults, life cycle of attack, defence strategies, and upcoming difficulties.\\
\end{abstract}

{\bf Keywords} 
\begin{enumerate}
    \item[] Security Attacks 
    \item[] Malware 
    \item[] Cybersecurity
    \item[] Ransomware attacks 
    \item[] Cyber Attacks
    \item[] Information Security
    \item[] Cryptocurrency\\
\end{enumerate}

\section{Introduction}

\noindent The infrastructure that is expanding the quickest right now is the Internet. Many contemporary technologies are altering the nature of human activities in today's technologically advanced society. Due to these new technologies, however, we are unable to fully safeguard our personal information, and cybercrime is on the rise. Brewer \cite{b1} claims that the phrases "ransom" and "malware" are the origin of the term "ransomware." It is a key element behind the surge of cyberattacks with the potential to make money off of victims. Noubir, conversely, thinks that while hackers previously found it difficult to benefit from assaults, this is no longer the case.  Cybercriminals are rapidly using ransomware assaults, or attackers who access a victim's data, encrypt it, and demand a payment \cite{a1}. \\

\noindent A form of virus called ransomware prevents users from using the attacked computer system. This sort of technical blackmail commonly involves drive-by assaults on maliciously designed web pages that take advantage of software and hardware flaws. It manifests as Manamecrypt, CryptoWall, CryptoDefense, or Cryptolocker \cite{b2}. Most of the files they targeted were in document storage formats like Office, PDF, and CSV, so they utilised strong encryption to scrambled virtually all of them, rendering them hard to retrieve without the special, secret key used to encrypt them. After receiving money, the cracker posts a display notice on the computer screen outlining the steps to do in order to retrieve the encrypted files, which puts an end to cryptovirology \cite{b3}.\\

\noindent Ransomware is more sophisticated and upgraded harmful software that takes the shape of Crypto or Locker and is designed to assault and retake control of computer systems and fundamental infrastructures. The great majority of these threats are intended to steal money from the victims either directly or indirectly by demanding a ransom in exchange for the decryption keys. In order to identify potential remedies, this systematic research searched through a variety of academic literature to examine the anatomy of ransomware, including its patterns and style of assaults \cite{b4}, \cite{a2}. The information technology infrastructure is seriously affected by the ransomware assaults. The effects of these assaults include system disruptions at most organisations, data loss attributable to file encryption, financial costs to the businesses for incident response and other security-related difficulties, and fatalities as a result of unplanned outages of some crucial medical equipment \cite{b5}. \\

\noindent In the literature, there have been previous systematic reviews of ransomware. The fundamental problem with those earlier evaluations is that they mostly focused on ransomware in the healthcare industry and other specialised areas, despite the fact that it is generally recognised that ransomware has no clear domain boundaries.
The full ransomware attack lifecycle and features presented in this paper can be used as a starting point for more study on ransomware. Additionally, current ransomware detection methods are discussed, along with the advantages and disadvantages of each method. Moreover, this paper will cover some discussion on prevention tools for ransomware attacks.

\subsection{Prior research}

\noindent Security lapses in computer networks can happen when a connection or network vulnerability is exploited to injure, undermine, or otherwise compromise the user. Active attacks and passive attacks are the two basic categories of attacks. These use a variety of techniques and procedures to steal information, identities, or money actively or passively \cite{b6}.\\

\begin{figure}[htbp]
\centerline{\includegraphics[width=\linewidth]{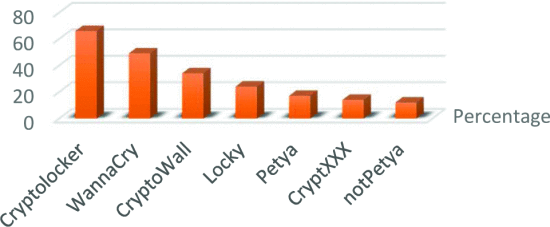}}
\label{fig:1}
\caption{List of Ransomware Attacks}
\end{figure}

\noindent In 1989, Joseph L. Popp created the original ransomware virus. Joseph is acknowledged as the father of ransomware as well \cite{b7}. A ransomware threat is frequently referred to as the Aids Trojan or PC Cyborg. Both corporate and human adversaries can use ransomware. It may spread to PCs via infected USB sticks or any type of attachment or link in phishing emails. The most recent Trojan horse to emerge is ransomware, a serious threat that has been progressively increasing in recent decades. The threat of ransomware often encrypts user data or deletes the necessary information while keeping the decryption key until the victim pays a ransom; this is largely due to bitcoins' untraceable characteristics. It can be vividly seen in the \hyperref[fig:1]{Figure 1}\\

\noindent Types of a Ransomware attack:
\begin{enumerate}
\item Crypto Ransomware - In a Crypto ransomware \cite{b8} attack, it encrypts our critical information on a device such that we are unable to decrypt it. Attackers of crypto ransomware make money by demanding a ransom before telling their victims that if they don't pay, their files would be lost forever.\\

\item Locker Ransomware - The Locker ransomware assault just locks the victims out of their operating system or laptop and prevents them from accessing it, rather than encrypting any data \cite{b9}. Attackers using locker ransomware demand a ransom in order to gain access to the laptop's operating system or other encrypted files.\\
\end{enumerate}

\begin{figure}[htbp]
\centerline{\includegraphics[width=\linewidth]{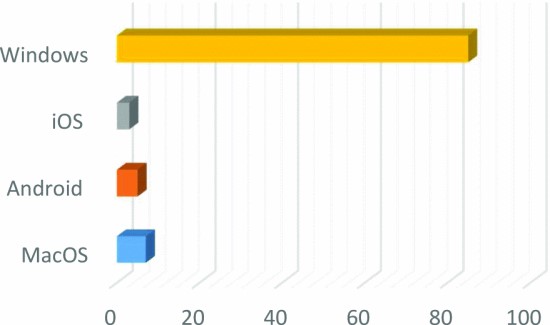}}
\label{fig:2}
\caption{Ransomware attacks on OS}
\end{figure}

\noindent Furthermore, operating systems play an inevitable role in ransomware assaults. We can certainly observe that devices running the Windows operating system are more susceptible to ransomware attacks and are frequently targeted \cite{b10},\cite{a3}. Additionally, other operating systems such as iOS and MacOS are also susceptible, proving that no OS is immune to the ransomware onslaught. This has been visualized in \hyperref[fig:2]{Figure 2}.\\

\noindent According to an existing literature that was conducted in the area, ransomware is recognised as a possible early 1990s exploitation of cryptography \cite{b11}. Even yet, it couldn't be used to demand money from people because it was simple to find the recipient of the money. The idea of using ransomware to make money was only possible until crypto-currency entered the scene. Thus, the invention of crypto-currency can be linked to the emergence of ransomware. \\

\begin{figure}[htbp]
\centerline{\includegraphics[width=\linewidth]{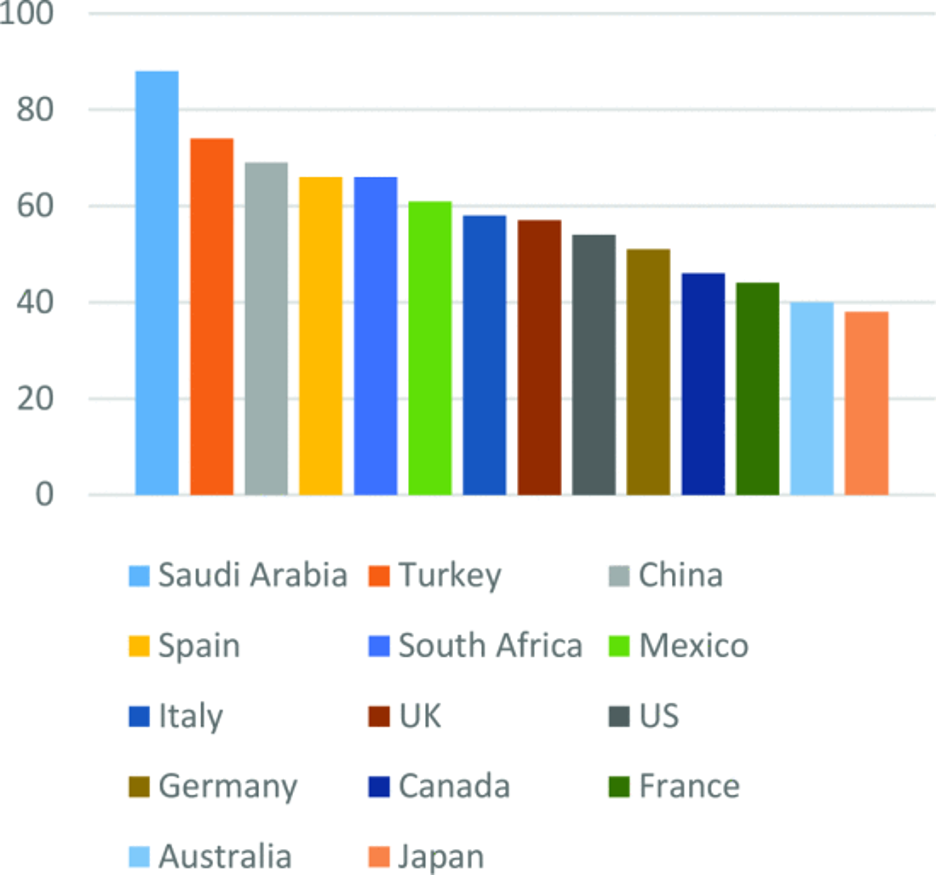}}
\label{fig:3}
\caption{Number of organizations that reported a ransomware attack in 2019 by country}
\end{figure}

\noindent Besides this, we also perform some critical analysis on the number of organizations are affected by ransomware attacks in different nations. Below figure shows the countries name where the greatest number of organization are negatively impacted by the ransomware in 2019, So ransomware affected about average of 50\% of organizations countries. It can be observed in \hyperref[fig:3]{Figure 3}.\\

\subsection{Research goals}

\noindent The goal of this study is to review previously conducted research, synthesise its conclusions, and focus on ransomware attacks critical analysis, threats, and prevention methods. We created three research questions to help concentrate the effort, as indicated in \autoref{tab:Table1}.\\

\subsection{Contributions and layout}

\noindent For individuals interested in ransomware attacks and cyber security to advance their work, this systematic literature review complements previous research and offers the research contributions in \autoref{tab:Table1}.\\

\begin{table}[htbp]
\caption{Research Questions}
\label{tab:Table1}
\setlength{\tabcolsep}{10pt} 
\renewcommand{\arraystretch}{1.5} 
\begin{tabular}{| p{10em} | p{5cm} |}
\hline
\textbf{Research Question} & \textbf{Discussion}\\ 
\hline
What are the different phases and variants of the ransomware attack in cyber security? & The evaluation of the ransomware attacks and the lifecycle of that attacks. A review on different variants of the ransomware attacks will help to perform the critical analysis and growth of the threaten attacks in information system \cite{a7,a10}. \\
\hline
What are the detection methodologies used to prevent the ransomware attacks? & Cybercriminal is being evolved day by day but to maintain the security on confidential data there are some research in field of detecting the ransomware attacks. The dynamic analysis process, file system, registry, and network activities are all investigated in this study in order to detect any loophole in systems.\\
\hline
 What methods and tools are available to prevent the ransomware attacks to manage security? & Ransomware is an active attack which can be prevented if the user follows and complies with security mechanisms. Researchers developed some prevention tools in order to protect the crucial data.\\
\hline
\end{tabular}
\end{table}

\begin{itemize}

\item We discover 27 relevant papers on cyber security and ransomware threats. This list of studies can be used by other academics to advance their respective work in this particular area.
\item We categorise ransomware attacks methods in a taxonomy.
\item We examined the criteria used to measure the defence, detection, and ransomware assault techniques and prevention tools for it.
\item We outlined the research data that was accessible for a later investigation of the anatomy of ransomware and produce guidelines to support further work in this area. 
\end{itemize}

\noindent The format of this paper is as follows: The techniques used to choose the primary studies for analysis in a methodical manner are described in Section 2. The results of all the primary research chosen are presented in Section 3. The findings in relation to the earlier-presented study topics are discussed in Section 4. The research is concluded in Section 5, which also makes some recommendations for more study.

\section{Methodology}
\noindent We carried out the Systematic literature review in order to accomplish the goal of responding to the research questions. This research review's objective is to present an overview of a systematic literature review on ransomware assaults, including the many kinds of ransomware attacks and mitigation strategies. An approach will outline how to look up different ransomware attack related articles, papers, books, and journals.\\

\subsection{Primary Studies}

\noindent The search capability of a specific search engine was emphasised by entering keywords. The keywords were chosen to encourage the development of study findings that would aid in addressing the study's issues. Only AND and OR were allowed to be used as Boolean operators. The query terms were: \\

\noindent (“ransomware” OR “ransom-ware” OR “Malware” OR “Type of ransomware attacks”) AND “security” \\

\noindent (“ransomware” OR “ransom-ware” OR “Malware AND (“cyber security” OR “cybersecurity” OR “cyber-security”) \\

\noindent The platforms used for this study were:
\begin{itemize}
\item[--] ABI/INFORM (ProQuest)
\item[--] Google Scholar 
\item[--] Web of Science
\end{itemize}

\noindent Depending on the search platforms, the title, keywords, or abstract were used in the searches. we have conducted the searches and processed all studies that had been published up to that point. The inclusion/exclusion criteria, which will be provided in Section 2.2, were used to filter the results of these searches. These iterations were performed both forward and backward until no further publications that met the inclusion criteria could be found.\\

\begin{figure}[htbp]
\centerline{\includegraphics[width=\linewidth]{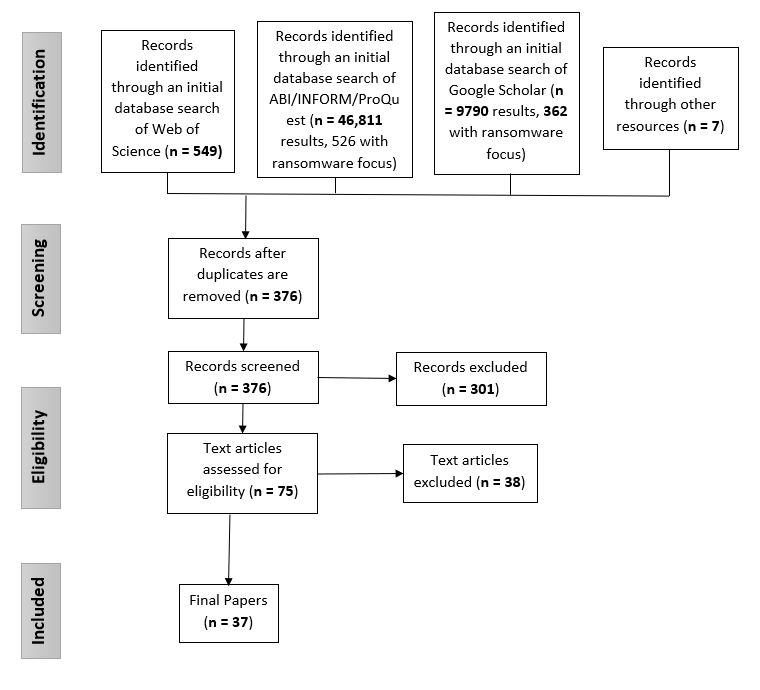}}
\label{fig:4}
\caption{Scoping Review Process}
\end{figure}

\noindent According to Chen and Bridges, ransomware is a category of self-replicating software that encrypts data before requesting a ransom. Review of unpublished theoretical tests and extensions that are currently being worked on, as well as an analysis of published theoretical literature. For the systematic study and analysis of the literature review, we will employ a three-step methodology. The diagram in Figure \hyperref[fig:4]{Figure 4} is used as a visual to display the library search and steps for finding articles on the specified topic.\\

\noindent In ProQuest, the "anywhere" search option was selected. The "anywhere in the article" search option in Google Scholar includes all authors, subjects, title information, abstracts, and keywords. The "subject" search parameter was used in the Web of Science. Journal articles, book sections or chapters, scholarly articles, working papers, conference papers, dissertations, and reports have all been included in this search.\\

\noindent Secondly, advanced filters have been implemented on the search which includes advanced filters like last 12 years, filter out source/document type and language specification to English. For advanced search, the new keywords "challenges and analysis" and "cyber danger" were utilised.\\

\noindent The articles have been reviewed based on slimming approach of the abstract and main body. After initial screening, all duplicates have been removed from the list.  The search field for all databases was restricted to English-language textual sources. 37 journal articles were chosen for the literature study after the final analysis. The procedure that resulted in the final 37 papers is shown in Figure \hyperref[fig:4]{Figure 4}.

\subsection{Inclusion and exclusion criteria}
\noindent Studies for this systematic literature review must present empirical findings and may include case study articles, new type of ransomware attacks, and commentary on how ransomware mitigation technologies may advance current security measures. They must be written in English and subjected to peer review. Google Scholar could return papers that are of a lower calibre, so all results will be examined for compliance with these standards. A study will only be included in this literature review if it has been updated in recent years. \autoref{tab:Table2} displays the main inclusion and exclusion standards.

\begin{table}[htbp]
\caption{Inclusion and exclusion criteria for primary studies}
\label{tab:Table2}
\setlength{\tabcolsep}{10pt} 
\renewcommand{\arraystretch}{1.5} 
\begin{tabular}{| p{10em} | p{5cm} |}
\hline
\textbf{Inclusion Criteria} & \textbf{Exclusion Criteria}\\ 
\hline
Practical information about ransomware assaults and other types of attacks must be presented in the article & papers examining how ransomware attacks affect business or the law \\
\hline
The paper must discuss ransomware or other relevant technology & Exclude governmental documents and blogs\\
\hline
The paper has to be a peer-reviewed document included in a journal or conference proceedings & non-English publications\\
\hline
\end{tabular}
\end{table}
    
\subsection{Selection results}
The initial keyword searches on the chosen platforms turned up a total of 549 studies. This was lowered to 376 when duplicate studies were eliminated. There were 75 publications left after the research were examined under the inclusion/exclusion criteria. The inclusion/exclusion criteria were again applied after reading all 75 papers, leaving 37 papers.\\

\subsection{Quality assessment}
In accordance with the guidelines, a review of the primary studies' quality was conducted. This made it possible to evaluate the papers' applicability to the research questions while taking into account any indications of bias in the research and the reliability of the experimental results. The evaluation procedure was modelled after certain literature reviews. The effectiveness of randomly chosen papers was examined using the following quality evaluation procedure:\\

\noindent \textbf {Stage 1:} Ransomware. The paper must be focused on various types of ransomware attacks or security breaches to a specific problem well-commented.\\ \\
\noindent \textbf {Stage 2:} Context. The objectives and conclusions of the research must be adequately contextualised. This will make it possible to understand the research correctly.\\ \\
\noindent \textbf {Stage 3:} Ransomware detection. The study must have sufficient information to accurately depict how the technology has been applied to detect attacks, which will help to address research questions.\\ \\
\noindent \textbf {Stage 4:} Security context. In an effort to aid in responding to research questions, the document must explain the security issue.\\ \\
\noindent \textbf {Stage 5:} Safety measures. Utilization of various safety measures to mitigate various types of ransomware attacks.\\ \\
\noindent \textbf {Stage 6:} Data retrieval. To assess accuracy, specifics regarding the data's collection, measurement, and reporting must be provided.\\

\noindent All additional selected primary studies were then subjected to this checklist for quality assessment. As demonstrated in \autoref{tab:Table3}, 10 studies were discovered to have failed to satisfy one or more of the checklist requirements and were thus eliminated from the Systematic literature review.\\

\begin{table}[htbp]
\caption{Excluded studies}
\label{tab:Table3}
\setlength{\tabcolsep}{10pt} 
\renewcommand{\arraystretch}{1.5} 
\begin{tabular}{| p{10em} | p{4cm} |}
\hline
\textbf{Checklist for the Criteria Stages} & \textbf{Excluded Studies}\\ 
\hline
\textbf{Stage 1:} Ransomware & \cite{S17} \cite{S10} \\
\hline
\textbf{Stage 2:} Context & \cite{S6} \\
\hline
\textbf{Stage 3:} Ransomware detection & \cite{S7} \cite{S12} \\
\hline
\textbf{Stage 4:} Security context & \cite{S20} \cite{S14} \cite{S11} \\
\hline
\textbf{Stage 5:} Safety measures & \cite{S2} \\
\hline
\textbf{Stage 6:} Data retrieval & \cite{S8} \cite{S16}\\
\hline
\end{tabular}
\end{table}

\subsection{Data extraction}
\noindent Data was then taken from all papers that had passed the quality evaluation in order to evaluate the completeness of the data and verify the accuracy of the information included within the articles. Before being extended to cover the entire set of research that have passed the quality evaluation step, the data extraction technique was first tested on a preliminary investigation. Each study's data was taken out, put into categories, and then entered into a spreadsheet. The following categories were applied to the data: \\

\noindent \textbf{Context data:} Information regarding the study's served objectives. \\

\noindent \textbf{Qualitative data:} The writers’ findings and judgements.\\

\noindent \textbf{Quantitative data:} Information gathered through trial and research have been used in the study.\\

\noindent The attrition rate of papers obtained from the first keyword searches on each platform down to the final selection of primary studies are shown in \hyperref[fig:4]{Figure 4}. along with the quantity of articles chosen at each stage of the process. \\

\begin{figure}[htbp]
\centerline{\includegraphics[width=\linewidth]{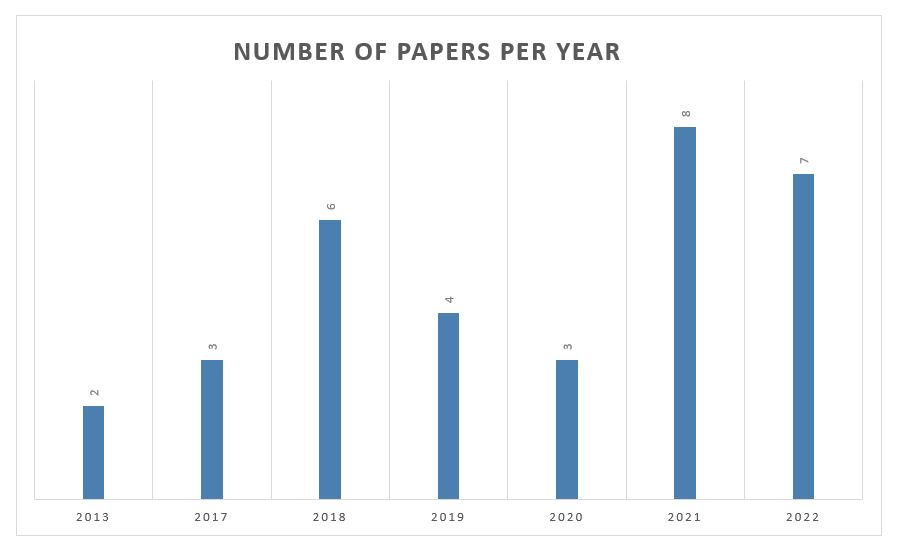}}
\label{fig:5}
\caption{The number of publications per year}
\end{figure}

\subsection{Data analysis}
\noindent We gathered the information contained in the qualitative and quantitative data categories in order to achieve the goal of responding to the study questions. We also performed a meta-analysis on the studies that were exposed to the last step of data extraction. According to the initial search results, this subject has important on a global scale. Numerous studies have been conducted in this field.\\

\noindent The majority of the articles are worthwhile, and value is referenced in the articles. Based on a set of requirements and criteria, the findings have produced value. The value has been described as a quantifiable result. "Ransomware attack mitigation" is among the significant benefits, while "Solution for ransomware attack" is among the impalpable advantages.\\

\subsubsection{Publications over time} ~\\
\noindent \hyperref[fig:5]{Figure 5} depicts number of articles that were published based on year. Among the 37 selected articles, there were (8) articles in 2021, (7) articles in 2022, (6) articles in 2018, (4) articles in 2019, (3) in 2017 and 2020, (2) articles in 2013. The findings demonstrate the topic's relevance and growing attention. As shown in figure, Research on ransomware attacks in relation to cyber security is on the rise. We anticipate that a sizable number of research articles on mitigation of ransomware attacks in practical applications will be published in the future.\\

\begin{figure}[htbp]
\centerline{\includegraphics[width=\linewidth]{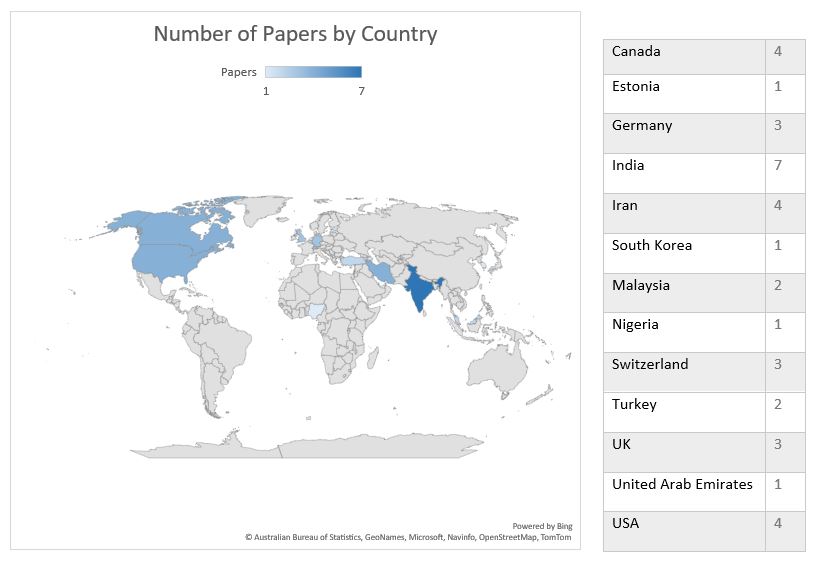}}
\label{fig:6}
\caption{Publications by country}
\end{figure}

\noindent \hyperref[fig:6]{Figure 6} displays the selected publication according to the location. Top locations include India (7), USA (4), Canada (4), Iran (4), Germany (3), Switzerland (3), UK (3), Malaysia (2), Turkey (2), Estonia (1), South Korea (1), Nigeria (1), and United Arab Emirates (1). Among all the chosen articles four of them were produced in more than one location.\\

\subsubsection{Significant keyword counts} ~\\
\noindent An analysis of keywords was done across all 37 studies in order to compile the common themes among the chosen primary research. \autoref{tab:Table4} lists the frequency with which various words appeared across all primary studies. As seen in the table, "Machine Learning" ranks third on of the most frequent term in our dataset, behind "ransomware" and "Trojan," after the author's chosen keywords, "ransomware" and "security." This demonstrates a growing interest in the ransomware attacks and security, which we shall cover in more detail in Section 3. \\

\begin{table}[htbp]
\caption{Keywords counts in the primary studies}
\label{tab:Table4}
\setlength{\tabcolsep}{10pt} 
\renewcommand{\arraystretch}{1.5} 
\begin{tabular}{| p{10em} | p{4cm} |}
\hline
\textbf{Keywords} & \textbf{Count}\\ 
\hline
Ransomware & 2312 \\
\hline
Trojan & 1673 \\
\hline
Security & 1564 \\
\hline
Machine learning & 1256 \\
\hline
Information & 753 \\
\hline
Deep learning & 655 \\
\hline
Software & 613 \\
\hline
Technology & 534 \\
\hline
Privacy & 498 \\
\hline
Attacks & 467 \\
\hline
Application & 440 \\
\hline
Crypto-currency & 425 \\
\hline
Malware & 370 \\
\hline
Analysis & 255 \\
\hline
Cybersecurity & 234 \\
\hline
IoT & 170 \\
\hline
\end{tabular}
\end{table}

\section{Findings}
\noindent After carefully reading each main research paper, pertinent qualitative and quantitative information was taken out and summarised in \autoref{tab:Table5}. Each of the primary studies had a goal or overarching theme related to earlier research on ransomware attacks. In \autoref{tab:Table5} below, the focus of each study is also listed.\\

\begin{figure}[htbp]
\centerline{\includegraphics[width=\linewidth]{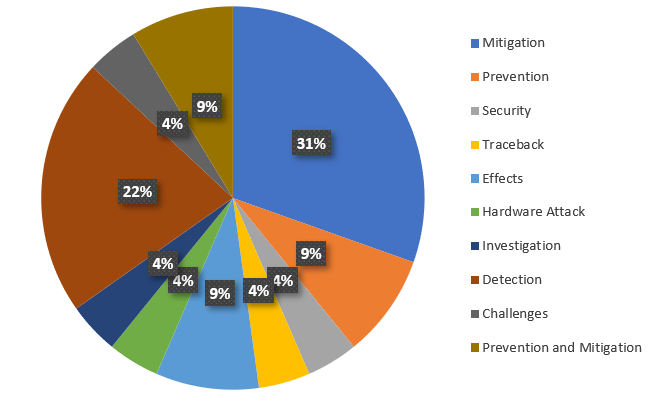}}
\label{fig:7}
\caption{Chart of Primary Research}
\end{figure}

\noindent To enable a more straightforward categorization of the issues of the primary research, the focus of each publication was further divided into wider groups.The research on ramming assaults that has been done in the realm of cyber security is shown in the table. Following table shows the study of 23 different research paper. \hyperref[fig:7]{Figure 7} shows the distribution of various research papers based on mail research area.\\

\begin{table}[htbp]
\caption{Main findings and themes of the primary studies}
\label{tab:Table5}
\setlength{\tabcolsep}{10pt} 
\renewcommand{\arraystretch}{1.5} 
\begin{tabular}{| p{4em} | p{4cm} | p{4em} |}
\hline
\textbf{Primary Study} & \textbf{Key Qualitative} & \textbf{Type of research}\\ 
\hline
\cite{S1} & By incorporating user-focused tactics, such as simulation and training on proper and thorough use of computers and network applications, health care companies may assure more reliable system defense \cite{a6,a4}. & Mitigation\\
\hline
\cite{S2} & It discusses Petya's ransomware's methodology and potential threats. This is also explored, as well as awareness and mitigation. & effects\\
\hline
\cite{S3} & The effects of ransomware attacks on users and providers of cloud services. In the paper, many mitigating tactics are covered. & Mitigation\\
\hline
\cite{S4} & By organising the literature into groups according to crucial factors, a taxonomy is created. A few reliable case studies are also provided to highlight how critically sensitive IoT devices are to attackers. & Challenge\\
\hline
\cite{S5} & The efficiency of clue analysis can be greatly increased by using the proposed strategy to trap the adversary in a deception environment. & Traceback\\
\hline
\cite{S6} & In order to provide a decryption key, the hackers who successfully encrypt user data want a ransom or money, frequently in the form of digital currencies. & Mitigation \\
\hline
\cite{S7} & Using the Volatility framework, memory forensics was used to analyze volatile memory dumps obtained from virtual machines. & Detection\\
\hline
\cite{S8} & In this report, NetConverse, a machine learning assessment study for reliable ransomware network traffic detection, is introduced. & Detection\\
\hline

\cite{S9} & Through checking, conglomeration, connection, and examination obligations, Situational Awareness(SA) empowers realizing what is happening in compromised gadgets and organization traffic. & Prevention\\
\hline

\cite{S10} & The suggested approach completely integrates with the ICE++ architecture, our earlier work, and uses Machine Learning (ML) techniques to identify and categorise the spreading stage of ransomware attacks that damage ICE. & Mitigation\\
\hline

\cite{S11} & Outlines the most recent security products, strategies, and research for preventing, mitigating, and containing ransomware outbreaks. & Prevention and Mitigation\\
\hline

\cite{S12} & Using machine learning to detect ransomware, this article suggests DNAact-Ran, a digital DNA sequencing engine. K-mer frequency vector and design restrictions for digital DNA sequencing are used by DNAact-Ran. & Detection\\
\hline

\end{tabular}
\end{table}

\begin{table}[htbp]
\ContinuedFloat
\caption{Main findings and themes of the primary studies(Continued)}
\label{tab:Table5}
\setlength{\tabcolsep}{10pt} 
\renewcommand{\arraystretch}{1.5} 
\begin{tabular}{| p{4em} | p{4cm} | p{4em} |}

\hline
\cite{S13} & This research presents a multi-tiered streaming analytics technique called a hybrid machine learner model that uses 24 static and dynamic attributes to classify different ransomware variants from 14 families. & Detection\\
\hline

\cite{S14} & How distinctive these patterns are and how they can be useful for intelligence exploitation to help stop ransomware assaults. & Investigation\\
\hline

\cite{S15} & The data of users should not be protected against attacks brought on by dangerous unidentified ransomware using signature-based malware detection technologies because they struggle to identify zero-day ransomware. & Detection\\
\hline

\cite{S16} & A quick overview of ransomware's history, the justifications for and against paying the demanded sum, and the best ways to avoid infection as well as recover from it should it occur. & Prevention and Mitigation \\
\hline

\cite{S17} & It outlines the significance of a written information security programme required by Massachusetts law or other information security framework. & security\\
\hline
\cite{S18} & Describe a methodology for categorizing ransomware attacks that ranks their virulence using a proposed classification algorithm based on file encryption and data deletion attack architecture. & Prevention\\
\hline

\cite{S19} & The Wade and seek approach is an adaptation of conventional hostage negotiation theory designed to assist businesses in minimizing the costs of ransomware outbreaks. & Mitigation\\
\hline

\cite{S20} & The main concept is to use Xen and DRAKVUF to intercept system calls in order to monitor for illicit activity directed at designated vital assets. & Mitigation\\
\hline

\cite{S21} & A hybrid method that checks network-based features, permissions, and text both statically and dynamically while keeping an eye on CPU, memory, and system call log utilisation. & Mitigation\\
\hline

\cite{S22} & The key contribution of this work is the attack, which is known as a hardware ransomware. Two hardware ransomware structures are shown as case studies, and a 65nm CMOS silicon demonstration is also provided. & Hardware Attack\\
\hline

\cite{S23} & This example of the Colonial Pipeline ransomware assault illustrates how reputational contagion can spread from firms directly touched by reputation events to organizations who are competitors of the firms. & Effects\\
\hline
\end{tabular}
\end{table}

\section{Discussion}
\noindent Locker and scareware were the first types of ransomware to spread. No data was encrypted by these malicious programmes. They only asked the victim to pay a fee to restore control after locking the computer's screen. In order to get the victim to pay a fine and have the charges against them dismissed, scareware would pretend to be a security agency that had captured the victim committing an offence.\\

\noindent 2018 is expected to see continued growth in the ransomware sector. The majority of IoT devices in use today still lack security, which is the cause of this\cite{b6}. Either there is no need for a password or other security measure to be in place since they are too basic to be secured by sophisticated passwords, or they are too complex to allow for such a measure. Lack of knowledge is a significant additional factor.\\

\subsection{RQ1: What are the different phases and variants of the ransomware attack in cyber security?}

\noindent Ransomware is characterized as having a life cycle that starts when malicious malware starts to propagate and ends when the victim is asked to pay a ransom. Knowledge of the ransomware life cycle can help you create a number of preventative measures that can successfully safeguard the target system's devices, files, and resources before the ransomware attack ever starts. Four stages can be used to examine the ransomware software life cycle.\\

\begin{enumerate}
\item Preparation phase- The ransomware life cycle starts after the prepared content created by the ransomware creator is served on the Tor network by a distributor. This stage establishes the procedure up until the ransomware is prepared for use. Sometimes, the attacker utilises the ransomware as is, other times they make a few minor adjustments to suit their needs.\\

\item Distribution phase- The dissemination stage is when ransomware is packaged as an email attachment or placed into a compromised website.\\

\item Phishing phase- An email or social media password, as well as credit card information, are obtained through a phishing assault. Using a bogus email that appears to be from a bank or other official entity, the victim is led to fraudulent or malicious websites. For their phishing assaults, attackers build phony web pages that mimic legitimate websites such as banking websites, social networking sites, email services, online games, and others \cite{b35}. On these pages, the victim's identification information, credit card number, password, and other details are sought. By paying attention to the requests in the email message and on the bogus website, the intention is to steal the victim's information \cite{a6,a4}.\\

\item Assault phase- In the ransomware attack stage, it starts to infect the victim's computer with a code dropper, mail attachment, or download from a compromised website. Several things start to happen once the ransomware has infiltrated the victim's system. The ransomware in the victim's computer performs tasks such as establishing a distinctive computer identity, blocking specific programmes, loading the programme to run automatically at startup, and stealing the internal IP address. Later, ransomware establishes a connection with the command and control (C\&C) server in order to obtain an encryption key. In the fourth stage, the malicious process looks for specific file extensions like pdf, doc, xls, pptx, and jpeg in the user's files \cite{b35,j4,j5}. After being moved to a different location, these files undergo a thorough encryption process. The encrypted files are renamed, and the old files are removed. The rogue machine then either sends the user a text file containing the ransom demands or shows them on their desktop. All of these processes are part of the life cycle of ransomware.\\

\end{enumerate}

\noindent With hundreds of new families emerging each year, crypto-ransomware has various variations. Here are some of the most well-known.

\begin{enumerate}
\item CRYPTOWALL- It belongs to a family of ransomware programmes that encrypt files and demand payment in early 2014 \cite{b6}. In subsequent versions, AES is used for encryption instead of the older RSA, which is then applied to a different unique public key. Thus, it becomes incredibly challenging to gain the key needed for file decryption.\\

\item CTB-LOCKER- On the user's hard disc, it is another another malware that encrypts data. A ransom demand is issued after decoding is complete. Elliptical curve cryptography is employed. It uses several languages, Tor, Bitcoin, and other technologies, and has fairly high infection rates \cite{b6}.\\

\item TORRENTLOCKER- This family of file-encrypting ransomware is only spread by spam email. It is locally targeted because both the original note and the ransom note are written in the regional tongue. Prior to demanding a Bitcoin ransom, it employs the AES algorithm. By obtaining email addresses from the victim machine, it extends its reach even further.\\

\item TESLACRYPT- It is among the newest ransomware. It encrypts specific files on the victim's PC using AES and then demands a ransom to unlock them.\\

\item CRYPVAULT- The ransomware in question is a straightforward batch script. Files are encrypted using RSA-1024 and given a new name with the suffix ".vault." \cite{b6}.\\

\end{enumerate}

\subsection{RQ2: What are the detection methodologies used to prevent ransomware attacks?}

\noindent It is crucial to remember that a number of strategies have been examined in the literature to reduce or completely eradicate the threat posed by ransomware. Both prevention and detection measures can be included under these actions. The best prevention tip for limiting the harm that ransomware might cause is to regularly backup your data. Even backups, however, can frequently get encrypted, and if you don't have the decryption key, decrypting information can be rather challenging \cite{a5,j1}.\\

\noindent A solution that can assist in preventing harmful encryption from occurring is ransomware detection \cite{j2,j3}. To combat this issue, numerous ransomware identifying techniques have been created.\\

\begin{enumerate}
\item Data Centric-Based Approaches- Data-centric identification tracks the sources that are impacted rather than the malicious activity that started the attack. Numerous studies have investigated data-centric crypto ransomware identification methods.\\
\noindent Entropy has been quantified in various studies to help identify ransomware outbreaks. To identify any alterations to a user's file structure, both before and after access, the suggested solution uses a statistical technique. Additionally, similarity assessment was used by the authors, which is fundamentally predicated on the idea that effective encryption produces a file that is entirely different from its original version.\\

\item Process-Centric-Based Approach- Process-centric-based detection keeps an eye on dangerous programmes' active processes for any unusual activity or behavior. Such actions could take the shape of specific, predetermined crypto ransomware events, including creating the encryption key or contacting specific APIs for applications that deal with cryptography (APIs) \cite{b37}.\\

\begin{itemize}
    \item Event-Based Detection- Event-based detection strategies look for certain (ad hoc) indicators that point to an oncoming crypto ransomware attack. Domain generating algorithms (DGAs), which generate new domains on demand, tracking command-and-control (C\& C) traffic to reveal any encoding set, and data transmitted between the infected virus and its remote C\& C server are all ways to detect ransomware before it starts its primary function \cite{b37}.\\
    
    \item Machine Learning-Based Detection- A number of research on the detection of crypto ransomware have also utilized machine learning approaches due to their effectiveness in virus identification. To simulate the behavior of crypto ransomware assaults, these researches use a variety of categorization techniques. Single-based and ensemble-based classifiers are two categories into which these classifiers can be divided. Individual machine learning classifiers are known as single-based classifiers, whereas ensemble-based classifiers aggregate different classifiers to perform complementary tasks on the same problem.\\

\end{itemize}

\item Delayed Detection- After watching the complete runtime data produced while the malicious programme was running, delayed detection occurs. This is because those detection models were developed using the whole runtime information for each malware case in the data set.\\

\item Early Detection- Early detection and forecasting aims to spot crypto ransomware threats before they start encrypting data. Early detection thus enables the implementation of proactive security measures prior to the start of the encryption process.\\

\end{enumerate}

\subsection{RQ3: What methods and tools are available to prevent ransomware attacks to manage security?}

\noindent The most dependable and effective way to safeguard personal information, data, and the entire computer system is to combine several tools—knowledge, software, and prudence. The greatest ransomware attack defenses should be set up in three different ways \cite{b38}.\\

\noindent Prevention steps to prevent ransomware attacks are as following:

\begin{itemize}
    \item Maintaining a current archive off-site and frequently backing up your data. Instead than relying on ransomware, use backups to protect your files. Check to see if the files being backed up are encrypted. Only you are able to recover it.\\
    
    \item Be cautious while opening links and attachments from unknown sources.\\
    
    \item Administrator, don't log in for any longer than is necessary. Give only what is necessary.\\
    
    \item Cybercriminals employ ransomware attacks to hack other users' accounts and add as many businesses as they can to them via fake links \cite{b7}.\\
    
    \item Create antivirus or anti-malware software and do frequent testing with it.\\
    
    \item Windows Firewall should always be fully configured and turned on. To limit the range of a potentially harmful gadget, firewall configured IP addresses \cite{b7}.\\
    
\end{itemize}

\noindent Tools to prevent ransomware attacks are as following:

\begin{enumerate}
\item Webroot SecureAnywhere- For consumers seeking a precise security tool that uses minimal resources while being successful against ransomware. Both individuals and small enterprises can use this technology effectively.\\

\item Trend Micro RansomBuster- By simply keeping important information and files in a protected folder and preventing any unwanted access to it, you may combat the issue by using this ransomware prevention application.\\

\item Malwarebytes Anti-ransomware- Uses behavior analysis to find harmful intent, which is something that no antivirus software can genuinely do.\\

\item SpinOne- Through ransomware defence, backup and recovery, risk analysis, compliance audits, and sensitive data security, this SaaS suite prevents data loss.\\
 
\end{enumerate}

\section{Future research directions of Ransomware attacks in cyber security}

\noindent The research on ransomware is still in progress, and it has numerous open issues. The research and survey articles analysed above demonstrate that there is presently minimal study on prediction models for ransomware because the majority of studies are centred on either detection or prevention. This is mainly to the fact that the majority of ransomware prevention and investigation approaches rely on data or information obtained during the execution of the assaults, making them useless given the characteristics of this new attack. Additionally, runtime data is used to get pre-encryption data from the ransomware's dangerous code. Pre-encryptions of this type do not protect against ransomware attacks.\\

\noindent Providing superior pre-diction techniques capable of foreseeing future ransomware is one of the most successful cyber security measures against it moving ahead. This should be accompanied by a tiered security framework. Consequently, ransomware attack forecasting models are more accurate efficient under certain circumstances. Attacks by ransomware are quite dynamic in nature. Therefore, early detection systems with ensemble classification principles and incremental learning are required to provide active and effective crypto ransomware detection and prevention systems in order to tackle the dynamic nature of crypto ransomware assault techniques. More distinctive and targeted characteristics of the malware must be designed for efficient ransomware assault detection. For the current detecting systems to attain a high degree of accuracy and precision, this is crucial. This is also directly tied to another largely unsolved issue with ransomware availability. Another obstacle to the quick development of detection and preventive systems is dataset.\\

\noindent The current ransomware payment systems rely on digital currency online payment systems like Bitcoin. These sites provide anonymous transactions, making it incredibly challenging to track the location of the ransom money. New techniques for cyber profiling must be created, and cryptanalytic transaction tracing procedures must be enhanced, in order to monitor and identify the attackers. Ransomware assaults will become less appealing to cybercriminals in this way. The likelihood of ransom recovery and data recovery during a ransomware assault will likewise rise as a result of these new, enhanced cryptanalysis approaches.\\

\noindent Researchers have been trying to develop a long-lasting defence against the ransomware attacks as a result of the damaging effects of the ransomware. In turn, the suggested methods are anticipated to significantly decrease its negative effects, avoid the attack, or maybe, if possible, eliminate it from existence. Analysis on ransomware activities, including its attack model, encryption techniques, use of Bitcoin, mode of operation, prevention, and protection mechanism, inundated the literature as a result of the need to find a solution to ransomware. Despite the enormous efforts made by experts, a long-lasting solution to ransomware has yet to eventuate.\\

\section{Conclusion}

\noindent In the present state of computer network security vulnerabilities are constantly being found and used. Every day, attacks take place that compromise data integrity and privacy. One variation of these attacks is the use of malicious software. Even though there are many various types of malicious software, they are all together referred to as ransomware because they demand a ransom from the victim. There are many different ransomware variations, all with different strengths, on the market. One of the most significant challenges in the digital world currently is ransomware. It is becoming a major danger and issue for businesses, organisations, healthcare systems, and information security specialists. Given the potential financial rewards, ransomware has attracted the attention of numerous cybercriminals, which has supported its explosive expansion. The majority of nations have been affected by ransomware. Regular precautions should be made to back up crucial data and files. The purpose of this paper is to emphasise the importance of learning about the ransomware attack and taking appropriate safeguards and procedures to get prepared for and reduce the risk from ransomware attacks. Moreover, we discuss ransomware and its various variants and evolution of ransomware. In addition to security measures, preventive and awareness campaigns may help prevent the ransomware's execution. In the future, we concentrate on creating a security system that thwarts ransomware attempts. There are ways to identify and resist other harmful attacks, but there is still no way to stop or even notice a ransomware attack. Computer infections can only be avoided by taking simple preventative measures. In these circumstances, user knowledge is crucial since social engineering is also used. Anti-ransomware software packages have been developed by several researchers, but their acceptability and practicality have not yet been determined.\\

\noindent \textbf{Declarations of interest} \\
\\None\\

\noindent \textbf{Acknowledgement} \\
\\We thank Professor Ali Dehghantanha (Ph.D.) for providing guidance and reference for the literature review. We also thank all the open-source resources and platforms such as, overleaf, Zotero and university of guelph library resources for providing the required resources and learning platform.\\

\renewcommand{\refname}{Primary Studies}

\end{document}